# Force Induced DNA Melting

Mogurampelly Santosh[†] and Prabal K Maiti[*]

Center for Condensed Matter Theory, Department of Physics,

Indian Institute of Science, Bangalore-12

**Abstract**

When pulled along the axis, double-strand DNA undergoes a large conformational change and elongates roughly twice its initial contour length at a pulling force about 70 pN. The transition to this highly overstretched form of DNA is very cooperative. Applying force perpendicular to the DNA axis (unzipping), double-strand DNA can also be separated into two single-stranded DNA which is a fundamental process in DNA replication. We study the DNA overstretching and unzipping transition using fully atomistic molecular dynamics (MD) simulations and argue that the conformational changes of double strand DNA associated with either of the above mentioned processes can be viewed as force induced DNA melting. As the force at one end of the DNA is increased the DNA start melting abruptly/smoothly after a critical force depending on the pulling direction. The critical force $f_m$, at which DNA melts completely decreases as the temperature of the system is increased. The melting force in case of unzipping is smaller compared to the melting force when the DNA is pulled along the helical axis. In the cases of melting through unzipping, the double-strand separation has jumps which correspond to the different energy minima arising due to different base pair sequence. The fraction of Watson-Crick base pair hydrogen bond breaking as a function of force does not show smooth and continuous behavior and consists of plateaus followed by sharp jumps.

---

[†] santosh@physics.iisc.ernet.in
[*] maiti@physics.iisc.ernet.in

1. **Introduction**

DNA melting is the process of breaking of Watson-Crick (WC) hydrogen bonds (HB) in double-stranded DNA (dsDNA) to form two separate single-stranded DNA (ssDNA). Thermodynamically stable dsDNA can be denatured by increasing the temperature or by applying force at one end of dsDNA or doing titration with acid/alkali [1]. In living organism strand separation can be induced by enzyme or proteins. The process of DNA strand separation resulting in the melting of dsDNA whether induced by increasing temperature or by force is fundamental in understanding important biological processes such as DNA transcription and replication which requires opening of the two strands of DNA. In the case of thermally assisted melting the unbinding of the two strands occurs with the increase in temperature [2] and the melting transition can be detected by calculating the fraction of HB ($f_{hb}$) breaks as a function of temperature. At low temperature all the WC base pairing remains intact and $f_{hb}$ is one. Thermal fluctuations can cause a small fraction of base pairs to loose their WC base pairing and forming transient denatured bubbles whose size can vary from few broken base pairs to 200 base pairs. These denatured bubbles can be monitored by single molecule experiments and understood in terms of stochastic approaches [3-6]. Bubbles in different regions of dsDNA can also coalesce with increasing temperature [7]. As the temperature is increased WC base pairing starts breaking and $f_{hb}$ keep on decreasing and finally goes to zero when the two strands get separated completely. So for homopolymer (DNA having only AT or GC base pairs) $f_{hb}$ decreases smoothly with temperature. However, for heterogeneous sequence variation of $f_{hb}$ as a function of temperature shows steps like behavior due to the difference in HB base pairing energy of AT vs GC. Depending on the base sequences the melting temperatures also vary accordingly: DNA having AT rich domains will melt at lower temperature compared to DNA having more GC rich domains. The process of thermally assisted DNA melting has been studied extensively over last few decades in the framework of Poland-Scheraga model [8-13] which was proposed more than 40 years ago and the model has been progressively refined to understand various aspects of DNA melting [14-16]. This model consists of alternate regions of denatured loops (single stranded) and bound segments (double stranded). The denatured loop regions are dominated by entropy gain on disruption of base pairs and the bound segments are dominated by the hydrozen bonding of base pairing as well as base stacking. Thus, double helical bound segments are energetically more favorable over the single stranded denatured loops whereas the single stranded denatured loops are entropically more favorable over double stranded segments. As the denatured/melting conditions are increased (such as force, temperature etc.) the loops start growing and finally at critical melting conditions the dsDNA separates into two ssDNA. It was found that the order of phase transition is determined by the critical loop exponent $c$ of the underlying loop class which determines the average loop size $\sim 1/l^c$, where $l$ is the length of denatured loop [9, 17, 18]. The entropy of the denatured loop is calculated by modeling them as ideal random walks and self-avoiding random walks which predicts a continuous denaturation transition in both two and three dimensions. The force induced melting transition also observed to be a continuous phase transition [17]. However, inclusion of excluded-volume interactions between denatured loops and the rest of the chain seems to drive the transition from continuous to discontinuous [9]. Another class of model originally proposed by Peyrard-Bishop (PB) [19, 20] has also extensively used to study the DNA denaturation transition [19, 21-25]. In this model, the bases in the two strands are allowed to move only in the HB direction connected by a Morse potential representing the HB whereas the bases in the same strand are coupled harmonically. In the framework of PB model, various groups have analyzed the statistical mechanics of the DNA denaturation transition using transfer integral technique and have calculated the inter-strand separation as a function of the temperature. This model allows the local melting of HBs and formation of denaturation bubbles. Later several groups have used PB model to study DNA unzipping process [12, 26-28] as well. All these studies have provided increased insight into various aspects of the DNA melting and unzipping but controversies remain regarding the order of this melting/unzipping transition and little is known about the kinetics and intermediate states during the melting/unzipping transitions. We expect molecular dynamics simulation to play a significant role giving molecular level understanding of various stages of melting/unzipping transitions.

With the advance of single molecule experimental techniques like optical tweezers or atomic force microscopy (AFM) it has now been possible to study the structural details of the single DNA (both dsDNA and ssDNA) under external force at varying physiological conditions. Several experimental and theoretical groups [17, 29-39] have studied structural transformation of DNA by applying force at one end of the dsDNA. When subjected to an external force dsDNA exhibits different force-extension regimes [30-32, 40]. For example, in the low - force regime, the elasticity of dsDNA is entropy dominated and the experimental force-extension data obtained in these experiments can be excellently described by the standard entropic worm-like chain model [29, 30, 41]. At large forces, the stacking potential can no longer stabilize the B-form configuration of dsDNA and the (optimally) stacked helical pattern is severely distorted [42], and therefore a structural transition from canonical B-form to a new overstretched conformation called S-DNA is observed. The structural modification of the DNA under pulling was also studied through molecular mechanics by Lavery and co-workers [31, 43, 44] and they proposed a structural transition from canonical B-form to a new overstretched conformation called S-DNA depending on the pulling protocol. If both the 3' ends are pulled, DNA unwinds upon stretching and DNA adopts a ladder like structure. On the other hand if both 5' ends are pulled, double helix structure is preserved and the structure is characterized by strong base-pair inclination and a narrower minor groove compared to the original B-DNA. These molecular modeling studies on this B-S transition of the DNA indicated that the DNA can be stretched to twice its initial length without losing the hydrogen bonding between the DNA bases. Later in a series of paper Bloomfield and co-workers argue that the overstretching transition at high force regime can be viewed as the force induced melting of the two strands of DNA [45-48] instead of viewing as a transition to a new form (so-called S-DNA) of DNA. On the basis of the existing experimental and simulation studies it has not been possible to conclusively validate or disapprove either of the viewpoints of the DNA overstretching transition. With the advent of faster computer and more realistic force field for DNA simulation, there were attempts to study the DNA overstretching transition at atomistic level and several recent studies [49-52] have given increasingly detailed molecular picture, energetic and role of entropy in the DNA overstretching transition. There also exist several single molecule experiments where one pulls apart both the strands of DNA in the direction perpendicular to the helix axis like a zipper by pulling the 3' and 5' terminal at the same end of DNA [33-38]. The experiments have been performed either at constant displacement ensemble or at constant force ensemble. Unzipping experiments for homo-polymer at constant force ensemble shows that with the increase of force hydrogen bonding of successive base pairing breaks continuously and DNA undergoes an unzipping transition once the applied force exceeds a critical threshold value. However, for hetero sequence DNA unzipping transition shows jumps in the number of hydrogen bond breaking corresponding to the energy barrier required to break various base pair sequence. Theoretical studies have established that the number of unpaired bases (broken HB) near the unzipping transition diverges much more strongly for hetero sequence of DNA than for homopolymer DNA.

To our knowledge apart from the works by Piana [50] and Harris et. al.[51] , there are no other microscopic studies which give a molecular level picture of the overstretching transition of the DNA under high force and provide with the confirmation if there is a melting transition associated with this overstretching transition. There also exist no molecular level studies of the DNA unzipping transition. Here we report large scale fully atomistic MD simulation of DNA stretching/unzipping under external force and demonstrate the force induced melting of the DNA duplex. Atomic level description of DNA melting and unzipping can provide insight into the several biological processes like DNA replication, RNA transcription and interaction of proteins that specifically bind to DNA. The rest of the paper is organized as follows: in section 2 we give the details of simulation methodology, section 3 gives the detail of the results from DNA overstretching and unzipping simulations. Finally in section 4 we give a summary of major results and conclude.

## 2. Simulation Details

All MD simulations reported in this paper used the AMBER9 software package [53] with the all-atom AMBER99 force field [54, 55]. A forcing routine has been added to AMBER9 to do the simulation at constant force and is available upon request. The electrostatic interactions were calculated with the Particle Mesh Ewald (PME) method [56, 57] using a cubic B-spline interpolation of order 4 and a $10^{-4}$ tolerance set for the direct space sum cutoff. A real space cut off of 9Å was used both for the electrostatic and van-der Waals interactions with a non-bond list update frequency of 10.

The starting structure for the DNA duplex with the sequence d(CGCGAATTCGCG) was built using nucgen module of AMBER suite of programs. Using the LEaP module in AMBER, the DNA structure was immersed in a water box using the TIP3P model for water. The box dimensions were chosen in order to ensure a 10Å solvation shell around the DNA structure in its fully extended/unzipped form when the DNA melts. In addition, some waters were replaced by Na+ counter ions to neutralize the negative charge on the phosphate backbone groups of the DNA structure. This gives system size comprising 9026 water molecules and 22 number of Na+ ions in a box of dimension 47x50x122 Å$^3$ for pulling along the DNA helical axis. This corresponds to 130 mM of Na+ ion concentration. For the unzipping case we have 14986 water molecules and 22 number of ions in a box of dimension 140x69x49 Å$^3$. The system was then subjected to the equilibration protocol outlined in our previous work [58, 59]. We have used periodic boundary conditions in all three directions during simulation. The external force was applied at one end on O3' and O5' atoms on each strand as shown in figure **1**. For forcing along helix axis, we kept one end of DNA fixed and applied force on the other end O3' and O5' atoms of DNA. On the other hand for the case where force is applied perpendicular to the helix axis, we kept one end of DNA free and applied force on the other end. Different force attachments such as O3' - O3', O5' - O5' and O3' - O5' at the two ends of DNA (or O3' - O5' at the same end) will result different conformational structures during pulling [44]. The external force started at 0 pN and increased linearly with time steps depending on the forcing rate till the DNA melts completely. The rate of forcing used for these studies was 0.0001 pN/fs. For comparison, we have also studied the melting process by pulling at higher rate of 0.001 pN/fs. It should be pointed out that a typical pulling rate in an AFM experiment is of the order of 10000 pN/s. So our forcing protocol is several orders of magnitude faster than that used in single molecule experiment. Hence the magnitude of force required for overstretching or unzipping will be larger compared to those observed experimentally. To understand temperature dependence of the force induced melting transition we have also done the pulling and unzipping simulations at the following temperatures: 300K, 312K, 325K and 350K.

## 3. Results and discussion
### 3.1 DNA pulling along helical axis

Figure **2** gives the force-extension curve for the duplex DNA at various temperatures. The force-extension curve consists of an entropic regime where, the extension of DNA beyond its contour length is negligible and this regime continues till 50 pN. This is followed by a highly nonlinear regime where DNA gets stretched almost 50% to 60% of its initial length with slow increase in force and this regime continues till 200 pN. Beyond this regime is the elastic regime where the DNA helical structure starts to deform and at the end of this elastic regime DNA structure transforms to a ladder like structure (so called S form of the DNA). The elastic regime continues till 500 pN. From the slope of this elastic regime we can get estimate of the stretch modulus of DNA which turns out to be 750 pN (corresponding to a salt concentration of 130 mM) and compares well with the available experimental values. From the temperature dependence of the force-extension curves we can estimate how the stretch modulus of dsDNA changes with temperature and will be the subject of future publication. With increase in temperature the magnitude of force at which DNA extension is double of its initial contour length decreases clearly indicating the DNA melting. Beyond the elastic regime is the overstretched structure of DNA which is followed by strand separation. Whether the ladder like structure is a melted state or another form of DNA is the topic of debate in last few years. To have closer look at this issue in figure **3** we show instantaneous snapshots of DNA structure at various pulling forces for simulation at 350K. As the force increases, at around 150 pN (2$^{nd}$ snapshots) we see the appearance of "holes" or bubbles region (similar to those observed by Harris et. al. [51]) where several WC base paring is disrupted. Generally hydrogen bond is represented as D-H⋯A where D is the donor and A is the acceptor which is bonded to D through H-atom. In case of DNA, D is N atom and A is either N or O atoms depending on AT and GC base pairing. When the distance between D and A atoms is less than 2.7 Å and the angle ∠DHA is greater than 130°, we say that the atom A is hydrogen bonded to atom D otherwise the HB is broken. With increase in force the DNA extension increases sharply and the ladder like structure (see snapshot at 290 pN) is obtained at a force of ~ 255 pN where the DNA extension is 85%. At this extension $f_{hb}$ = 0.43 which means that almost 60% of the WC base paring is lost in the ladder like structure and so it is a partially melted structure. This might indicate that indeed the so called S-DNA is partially melted form of DNA. Of course the complete melting of the duplex (which we define to be the case when the number of broken HBs is 80% of initial HBs (i.e., $f_{hb}$ = 0.2) and the corresponding force as melting force ($f_m$)) happens at a force $f_m$ = 266 pN which is higher than the force required for the appearance of the ladder like structure. In general the ladder like structure is expected to obtain 10-15 pN below the melting force $f_m$. Note that similar picture emerges at other temperatures as well. It is also worth mentioning that at low forcing rate the above mentioned regimes in the force-extension curves shifts to much lower force values as will be discussed in section 3.2.

To see whether the force induced structural transformation is related to the melting of the dsDNA, or just gives rise to another form of DNA (so called S-DNA) we estimate the fraction of hydrogen bonded base pairs ($f_{hb}$) as a function of applied force for different temperatures. In figure **4** we plot $f_{hb}$ as a function of applied force at four different temperatures. We see that up to 108 pN all the WC base pairing remain intact leading to $f_{hb}$ = 1 for all temperatures. Once a critical force is reached, hydrogen bonds involved in WC base pairing starts breaking and $f_{hb}$ starts decreasing with increasing force. The magnitude of the critical force needed to initiate the HB breaking at 300 K is 162 pN which decreased to 108 pN when the temperature is increased to 350 K. It can be observed from figure **4,** that the melting force $f_m$, decreases with increased temperature. For example, the melting force $f_m$ is 416 pN at 300 K which came down to 266 pN at 350 K. Beyond the temperature dependent critical force, $f_{hb}$ decreases very sharply for all temperatures. However, the decrease is not smooth but shows steps like behavior separated by sharp jumps. This behavior can be related to the base pair sequence in the DNA because of the fact that for a given temperature AT rich domains melt at smaller force compared to GC rich domains.

To get more detailed microscopic view of the thermodynamic stability of the duplex DNA under stretching force we compute the internal energy of DNA as a function of its extension. To compute the internal energy of DNA we first partition the potential energy into a sum over atoms. This is done by assigning half the energy for every two-body interaction to each of the two atoms, all the energy for each three-body interaction and each four-body inversion term to the central atom, half the energy for every four-body dihedral (torsion) interaction to each of the two central atoms. Then we collect these atomic energies together for whole of the DNA. Thus, each atomic energy contains the interactions of that atom with the rest of the system. It also includes the solvation effects as the interaction energy term for each of the atom includes the contribution from the water as well as counterions. However, this energy at a given force does not include conformational fluctuations as it was not averaged over the canonical ensemble of structures. Energy variation as a function of force during pulling is shown in figure **5** at two temperatures. In the small force region of about 0-50 pN, there is very little change in the extension of the DNA, during which the conformational entropy dominates. During this period the water molecules and ions get reorganized around the surface of DNA resulting a better solvation. In this force region, the internal energy of the DNA decreases compared to its value in the zero force limit implying a thermodynamically more stable state at this extension. In steered MD simulation of DNA duplex stretching, Harris et. al.[51] also observed that DNA at an extension > 2 nm becomes more stable than the unrestrained double helix. With further increase in force, the DNA extension increases and internal energy of the DNA increases with extension leading to unstable DNA structure and eventually

DNA melts when a critical force $f_m$ is reached. At high temperature the increase in internal energy of DNA with extension is much higher and consequently DNA melts at smaller force.

**3.2 Effect of forcing rate**

In AFM experiment typical pulling rate ranges from 100 to 10000 pN/s. In our simulation the slowest pulling rate we have achieved is $10^{11}$ pN/s (or 0.0001 pN/fs) due to computation limitations. Depending on the rate of pulling the molecular adhesion bond strength varies. Theoretically it has been found that the bond strength increases as the logarithm of the pulling rate [60]. Therefore, when DNA is pulled at faster rate, the HB strength of the base pair is expected to increase dynamically and hence the DNA should melt at higher force. To see this feature, we have done the pulling simulation of DNA at a higher forcing rate of 0.001 pN/fs and compare the results with the pulling rate of 0.0001 pN/fs. The variation of $f_{hb}$ as a function of force at two rates is shown in figure **6**. We see that when DNA is pulled at faster rate, the HBs start breaking at larger force compared with slower rate of pulling. When increased the force after HBs start breaking, there is a sudden decrease in $f_{hb}$ and eventually goes to 0.2 where the DNA is in a melting condition. The corresponding melting force $f_m$ is very high when pulled with 0.001 pN/fs rate. It would be interesting to study the melting process associated with the overstretching transition when DNA is pulled with slower rate closer to the AFM pulling rate which will require significant computational resources.

**3.3 Unzipping Transition**

So far we have discussed duplex melting when the force is applied along the helical axis of DNA. When the force is applied perpendicular to the helical axis as shown in figure 1 (**b**), the two strands start separating from each other like a zipper and DNA undergoes an unzipping transition. We define the distance between 3' and 5' atoms at the same end where force is applied as ds-separation ($x$). The ds-separation is plotted as a function of force in figure **7** at various temperatures. Up to a critical force which depends on temperature there is no change in the ds-separation. Once a critical force is reached HB between the base pairs gets disrupted and two strands start separating from each other. The magnitude of critical force at which the two strands start separating from each other decreases with increase in temperature. This is due to the fact that the increased temperature helps overcome the free energy barrier of base pair opening and hence makes it easier to pull apart the two strands. For example at 300 K, ds-separation starts increasing at a critical force of 237 pN which is decreased to 131 pN at temperature 350 K. Beyond this critical force the ds-separation increases rapidly and shows jumps and pauses. Pauses and jumps are due to the large force that is required to overcome the energy barrier due to hydrogen bonding between base pairs. The pause duration (or width) decreases with increase in force and for very large force there is no pauses near the transition. Since DNA has double helical structure the torsional relaxation can also play significant role during the melting process. Pauses and jumps can occur if the breaking of hydrogen bonds happens much faster than the time scale of torsional relaxation. The magnitude of force during which an intermediate pause continues before another jump occurs strongly depends on the sequence. Near the melting transition the ds-separation grows very rapidly and after breaking of all HB's the two stands separate from each other. DNA unzipping experiments at constant force [38, 61] observed jumps in the ds-separation. Theory [24, 39] also predicts such jumps in ds-separation which are in very good agreement with our simulation results.

In figure **8** we plot the fraction of HB $f_{hb}$ as a function of force applied to unzip DNA at different temperatures. HBs start breaking at the critical force and $f_{hb}$ rapidly decrease with increased force. Over 120 pN, there were no broken bonds for all temperatures and eventually breaking was initiated beyond this critical force. Here also we use the same criteria for melting like in overstretching case i.e., DNA is melted when 80% HBs were broken and correspondingly the melting force $f_m$. The fraction of HB $f_{hb}$ as a function of force curve has also jumps and pauses which can be again attributed to the sequence effects. The melting of DNA observed at a melting force of 355 pN for temperature 300 K and at 253 pN for temperature 350 K clearly indicating that the melting force $f_m$ decrease with increased temperature. Figure **9** shows the instantaneous snapshots at various forces while unzipping the DNA. In the case of pulling along the helix axis, the AT rich region melts early at smaller forces compared to the GC rich region as shown in figure **3**. This is because the AT base pair contains two HBs whereas the GC base pair contains three HBs which require more force to break. But during unzipping we don't see the early breaking of AT base pair instead melting starts right from the end of the DNA where the force is applied.

## 4. Summary and Conclusion

To summarize, we use fully atomistic simulation to study the process of DNA melting under external force for short DNA duplex. When DNA is pulled along the helical axis at constant force it undergoes large conformational change and after a critical force (e.g. 100 pN at 300K) is reached DNA has an extension of 20-25% but no base pairs are melted. In the force range of 100-200 pN DNA extension goes up to 50%. During this extension we see the local melting of some base pairs. The snapshots at in the figure **3** indicate that the central AT base pairs melt earlier than the terminal GC base pairs during this extension. At an extension close to double its initial contour length DNA undergoes an overstretching transition. This transition can be viewed as a melting transition when analyzed in terms of the breaking of hydrogen bonds between the bases. Such picture is consistent with the earlier atomistic simulation studies by Piana [50] as well as experimental data from Bloomfield and co-workers [45-47, 62]. The value of $f_{hb}$ = 0 at certain force means that there is no hydrogen bonding between the bases of the opposite strands indicating the complete unbinding/melting transition of the two strands. The transition is highly cooperative in the sense that after a critical force $f_{hb}$ shows sharp decrease as a function of force with intermediate plateaus. Whether the presence of plateaus in the variation of $f_{hb}$ as a function of force indicates a discontinuous transition is not yet clear and needs further investigation. Also the presence of multi step patterns was not observed in earlier simulation studies [50, 52]. The force at which the unbinding transition occurs depends strongly on the temperatures. With the increase in temperature the free energy barrier to the melting decreases and DNA melts at lower force. The temperature dependence of the free energy of melting will be investigated in future. The force corresponding to the duplex melting at a given temperature strongly depends on the direction of the applied force as well. So for the unzipping case when the force is applied perpendicular to the helical axis, DNA melts at a lower force. Again this observation is consistent with the available experimental and theoretical observation [63]. During unzipping also we observe series of jumps and plateaus as is evident both in the DNA separation as well as in the variation of $f_{hb}$ as a function of force which corresponds to the energy barrier to break the different base pairing. These findings are in excellent agreement with the available literature data [38, 39, 61] and demonstrate that nanosecond long all atom simulations with the present DNA force field can give valuable microscopic details of the melting phenomena. Future study will focus on the DNA unzipping transition at various salt concentration, different lengths and base pair sequence of DNA and with different pulling rates. Another important aspect would to study the low temperature, low force region of the force-temperature (f-T plane) phase diagram to explore the re-entrant behavior in the DNA melting if any. Several theoretical studies have predicted a re-entrant region in f-T phase diagram at low T for finite forces where DNA denaturation occurs with decreasing T [17, 64, 65]. However, no microscopic studies or experiments exist to confirm or disapprove such claim.

**Acknowledgement**

We acknowledge the financial support from Department of Science and Technology (DST) and University Grants Commission (UGC), India. MS thanks Geetanjali for helping in drawing figure **1**.

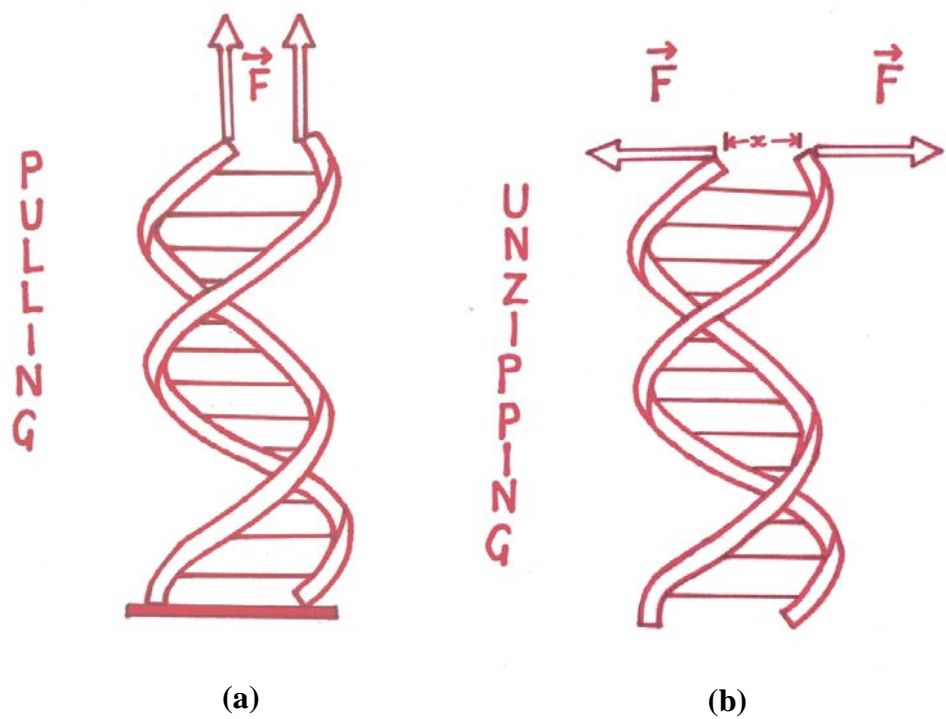

**(a)** **(b)**

Figure **1:** Schematic diagram of the forcing protocol for applying force (**a**) along the helix axis with a fixed end and (**b**) perpendicular to the helix axis with free end of DNA.

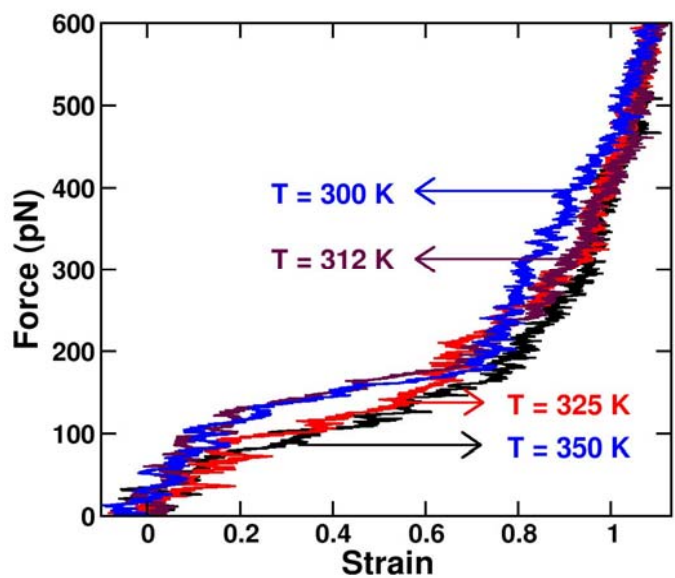

Figure **2**: Force-extension curve of 12-mer dsDNA when force is applied along the helix axis with 0.0001 pN/fs rate at various temperatures. With increased temperature the magnitude of force at which DNA extends double of its length, decreases indicating the force induced DNA melting.

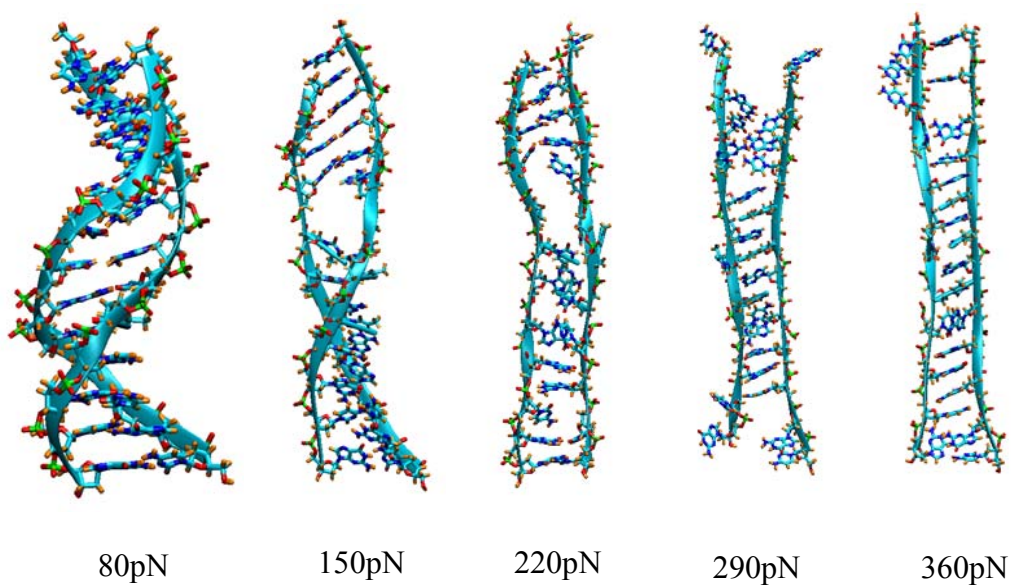

| 80pN | 150pN | 220pN | 290pN | 360pN |

Figure **3**: Instantaneous snapshots of DNA at various pulling forces when pulled along the helical axis. These snapshots correspond to T = 350K. Note that at 150pN when the DNA extension is around 25-30% AT base pairs in the middle of the DNA starts melting whereas terminal GC base pairs melt at higher extension occurring at higher force. Similar picture holds good at other temperatures as well. For clarity water molecules and ions were not shown in the picture. These snapshots have been generated using VMD software [66] developed at UIUC.

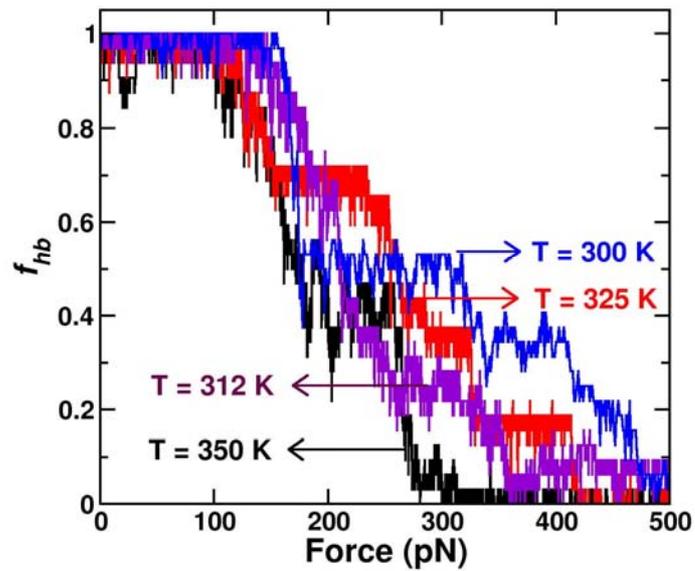

Figure **4**: Fraction of Watson-Crick H-bonds ($f_{hb}$) as a function of force applied along the dsDNA helix axis at various temperatures. The rate of pulling DNA is 0.0001 pN/fs. DNA melts at smaller force with increased temperature.

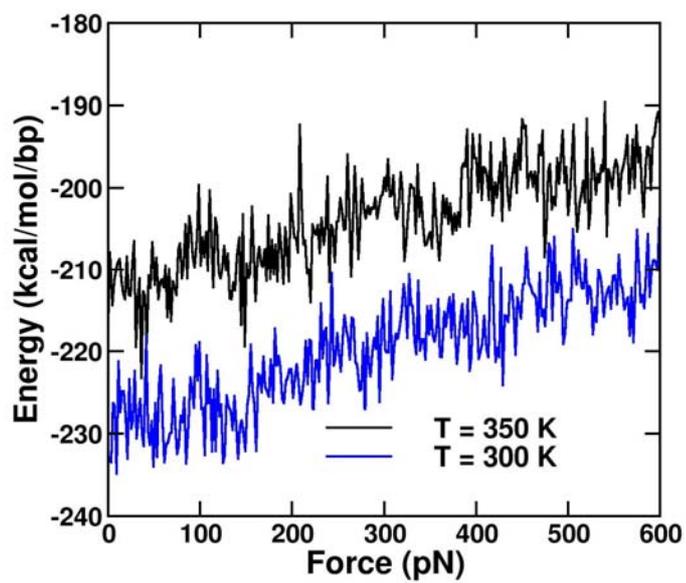

Figure **5**: Total internal energy of the DNA as a function of force at 300K and 350K. Increase of energy with pulling force indicates the destabilization of the DNA.

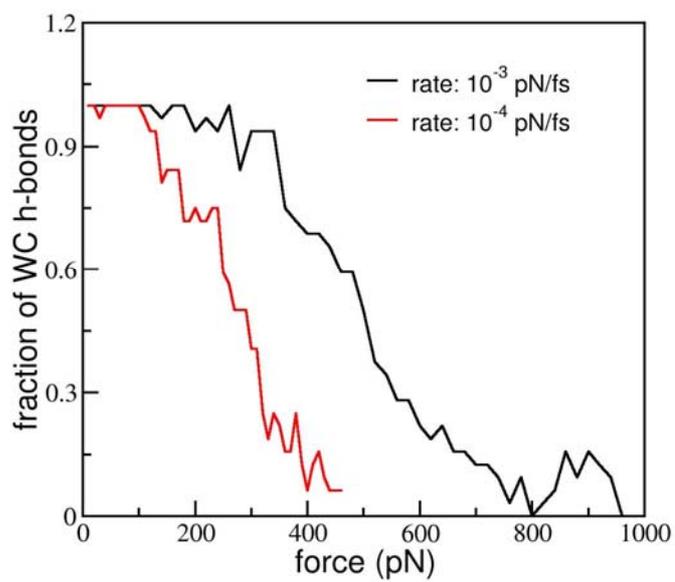

Figure **6**: Fraction of hydrogen bonds as a function of pulling force at two different pulling rates. With higher pulling rate of 0.001 pN/fs DNA melts at higher force compared to the case when DNA is pulled slowly at 0.0001 pN/fs rate.

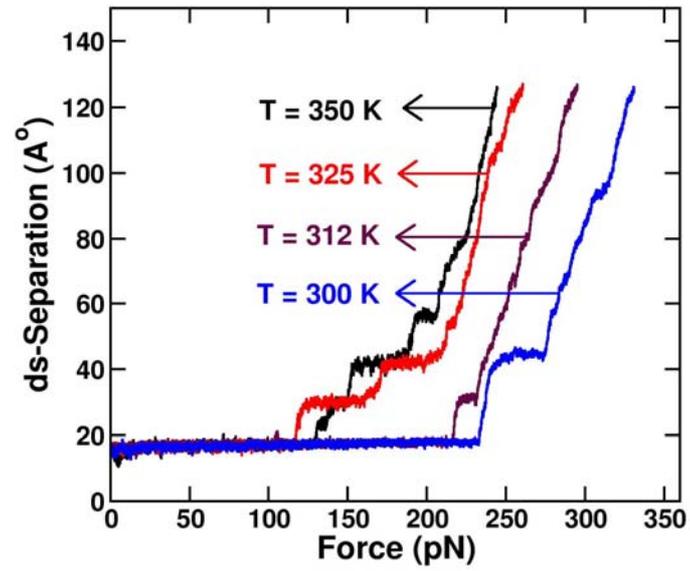

Figure **7**: ds-Separation of DNA as a function of unzipping force at various temperatures. We see jumps in the separation distance with unzipping force.

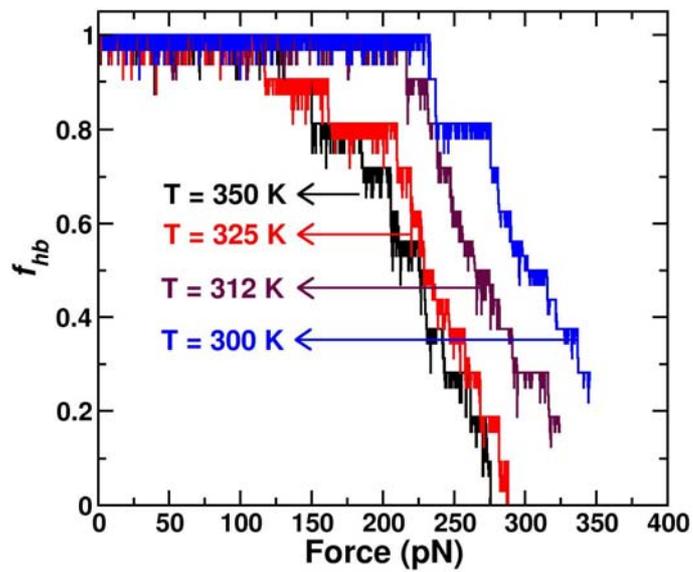

Figure **8**: Fraction of Watson-Crick H-bonds $f_{hb}$ as a function of force applied perpendicular to the dsDNA helix axis at various temperatures. The rate of pulling DNA is 0.0001 pN/fs. DNA melts at smaller force with increased temperature. We observe the jumps in the melting curve.

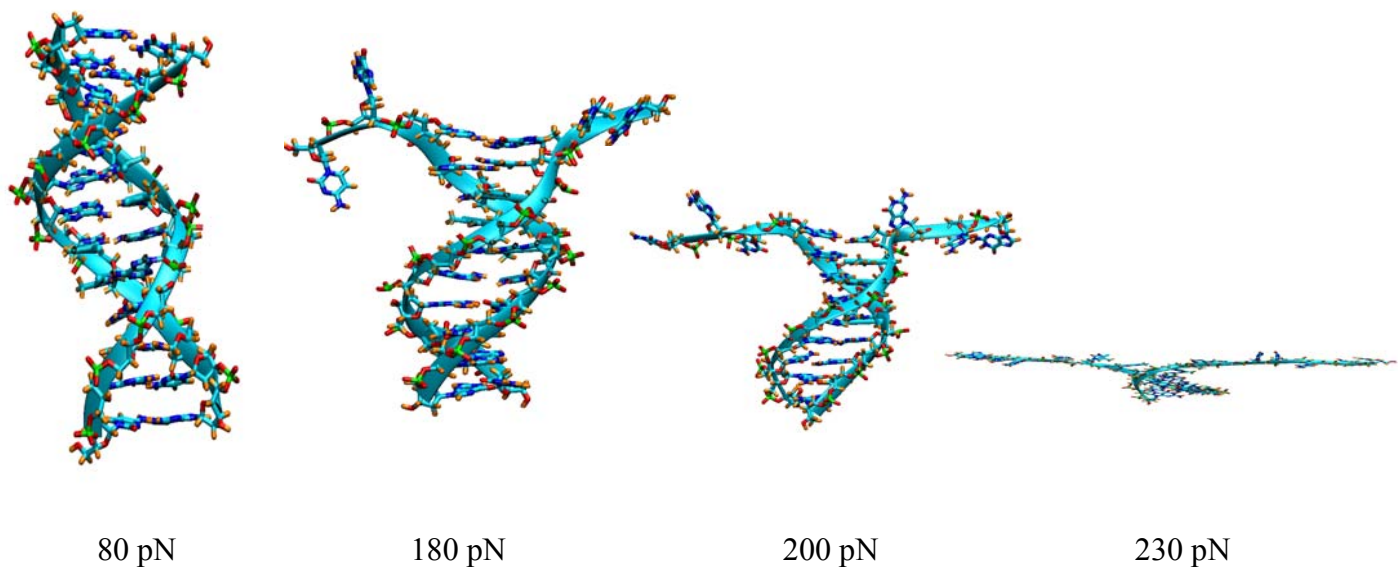

| 80 pN | 180 pN | 200 pN | 230 pN |

Figure **9**: Instantaneous snapshots of DNA at various pulling forces during unzipping. These snapshots correspond to T = 350 K. Unlike the pulling case along helical axis where the AT region melts early (figure 3), the melting starts from the end terminus where the force is applied. At a force of 253 pN, the DNA melts completely and further increment in force caused the separation of two strands from intact dsDNA. For clarity water molecules and ions were not shown in the picture. These snapshots have been generated using VMD software [66] developed at UIUC